\documentclass[12pt]{article}
\usepackage{comment}
\usepackage{amssymb,amsmath,comment,mathtools}
\usepackage[dvipdfmx]{graphicx}
\usepackage{subfigmat}
\usepackage{braket}
\usepackage{hyperref} 	  		
\usepackage{threeparttable}
\usepackage{epsfig}
\usepackage{here}
\usepackage{color}
\usepackage{bm}
\graphicspath{{./Figures/}}

\newcommand{\ba}{\begin{align}}
\newcommand{\ea}{\end{align}}

\def\bea{\begin{eqnarray}}
\def\eea{\end{eqnarray}}
 \makeatletter
\def\alt{\mathrel{\mathpalette\gl@align<}}
\def\agt{\mathrel{\mathpalette\gl@align>}}
\def\gl@align#1#2{\lower.6ex\vbox{\baselineskip\z@skip\lineskip\z@
\ialign{$\m@th#1\hfil##\hfil$\crcr#2\crcr\sim\crcr}}} \makeatother
\renewcommand{\thefootnote}{\fnsymbol{footnote}}

\begin{document}
\begin{flushright}
\end{flushright}
\rightline{NITEP 217}

\vspace*{1.0cm}

\begin{center}

{\Large\bf 
Indirect Detection for Higgs Portal Majorana Fermionic Dark Matter
}
\vspace{1cm}

{\large 
Naoyuki Haba${}^{a,b}$, Junpei Ikemoto${}^{a,c}$,  Yasuhiro Shimizu${}^{a,b}$ 
\\
and Toshifumi Yamada${}^{d}$
} \vspace{.5cm}

{\baselineskip 20pt \it
${}^{a}$Department of Physics, Osaka Metropolitan University, Osaka 558-8585, Japan \\
${}^{b}$Nambu Yoichiro Institute of Theoretical and Experimental Physics (NITEP),
Osaka Metropolitan University, Osaka 558-8585, Japan\\
${}^{c}$Institute of Science and Engineering, Shimane University, Matsue 690-8504, Japan\\
${}^{d}$Institute for Mathematical Informatics, Meiji Gakuin University, Yokohama 244-8539, Japan
}

\vspace{.5cm}

\vspace{1.5cm} {\bf Abstract} \end{center}

We study the $\gamma$-ray signal emitted from dark matter (DM) pair annihilation 
 in the Higgs portal Majorana fermion DM model.
In the model, a Majorana fermion DM $\chi$ couples with the Standard Model (SM) Higgs field $H$
 through a higher-dimensional term $-{\cal L}\supset H^\dagger H \bar{\chi}\chi/\Lambda$, where $\Lambda$ is a cutoff scale.
The pair annihilation of $\chi$ through the above term produces the Higgs boson and the longitudinal modes of $W,Z$ gauge bosons.
The Milky Way dwarf spheroidal satellite galaxies (dSphs) are used as the most promising targets to search for 
 the $\gamma$-ray signal of the model,
 due to high DM density and lack of astrophysical backgrounds.
The {\it Fermi} Large Area Telescope ({\it Fermi}-LAT) is used for the search, for its high sensitivity.
In this work, we use 14-year {\it Fermi}-LAT data from 16 dSphs, to constrain the DM pair annihilation cross section for the DM mass range from 125 GeV to 100 TeV.

\thispagestyle{empty}

\newpage
\renewcommand{\thefootnote}{\arabic{footnote}}
\setcounter{footnote}{0}
\baselineskip 18pt


\section{Introduction}
The existence of dark matter (DM) has been firmly supported by astrophysical and cosmological observations.
The relic abundance of DM in the present universe is determined \cite{pdg Astrophysic}:
\bea\label{eq:observedValue}
	\Omega_{\rm DM}h^2 = 0.12\pm0.0012
	\quad.
\eea 
As for the mechanism for generating DM, both freeze-in and freeze-out mechanisms have been widely studied~\cite{relic_density,Freeze-in/out}.
Also, various types of the interaction between DM and Standard Model (SM) particles have been proposed

In this paper, we consider the Higgs-portal Majorana fermion DM scenario,
 in which the Majorana fermion DM ($\chi$) and the SM Higgs boson (H) interact through a higher-dimensional operator,
 $-\mathcal L \supset H^\dagger H \bar \chi \chi / \Lambda$, 
 where $\Lambda$ is a cutoff scale~\cite{previous study}.
Here $\Lambda$ is so large that the DM is never in thermal equilibrium with the thermal bath of SM particles.
Still, this model can explain the relic abundance of DM Eq.~(\ref{eq:observedValue})
 through the freeze-in mechanism for generating DM~\cite{previous study}.

In our previous work Ref.~\cite{previous study}, we have investigated 
 direct detection of DM in the Higgs-portal Majorana fermion DM scenario,
 where the DM $\chi$ scatters off atomic nuclei in a detector~\cite{DM_DirectDetection,TASI_directDetection}.
In this paper, we study indirect detection of DM~\cite{Tasi_indirect,DM_indirect}, where the DM $\chi$ goes into pair annihilation creates 
 stable particles such as electrons, positrons, (anti-)protons, (anti-)neutrinos and photons.
In particular, we focus on the photon signal from the DM pair annihilation, because photon has no electric charge and so its flux is not affected by
 the galactic magnetic field. Also, photon is much easier to detect than neutrino.
 
Gamma rays originated from DM pair annihilation can be detected by the Large Area Telescope (LAT) in the 
 {\it Fermi Gamma-ray Space Telescope (Fermi) }
 \cite{Fermi-LAT2009}.
Such gamma rays are most likely to be produced in regions of high DM density.
Dwarf spheroidal satellite galaxies (dSphs) in the Milky Way are promising targets 
 due to their high DM density and low baryonic contamination.
Early studies 
 \cite{previous-LAT-study1, previous-LAT-study2,main-referred-study, main-referred-study2}
 set constraints on the DM annihilation cross section from the stacked LAT data of dSphs.
In this work,
 we use the likelihood functions of the previous studies
  and set constraints on the DM pair annihilation cross section in our Higgs-portal Majorana fermion DM scenario.

This paper is organized as follows. 
In Section 2,
 we describe the 14-year {\it Fermi}-LAT dataset we use.
In Section 3,
 we review the Higgs portal Majorana fermion DM model and explain the $\gamma$-ray flux from the DM pair annihilation.
In Section 4,
 we describe the DM distribution in the dSphs of the Milky Way that we adopt. 
In Section 5, we perform the combined likelihood analysis to derive a constraint on the DM pair annihilation.
In Section 6, we present the result of the analysis.
Section 7 concludes the paper.


\section{Data Setting}
\label{sec:dataSelect}

We examine 14-year {\it Fermi}-LAT data 
 (2008-8-14 to 2022-8-2)
 setting \texttt{P8R3\_SOURCE\_V3} class events in the energy range from 100 MeV to 1 TeV. 
The minimum energy, 100 GeV, is selected in order to reduce the effects of outflow from the bright limb of the Earth.  
In addition,
 to avoid contamination from terrestrial gamma rays,
 we reject events with zenith angles larger than $90^\circ$.
We use an open-source \texttt{fermipy} (Ver 1.2.2)
\cite{fermipy}
to analyze the {\it Fermi}-LAT dataset,
 which is built on the {\it Fermi science tools}
 \footnote{
\texttt{Fermitools} (Ver 2.2.0): 
\url{https://fermi.gsfc.nasa.gov/ssc/data/analysis/software/}
}.
We apply the recommended quality filter \texttt{DATAQUAL>0\&\&LATCONFIG==1} to ensure that the data are valid for scientific analysis.
This filter selection takes into account the loss of observing time due to the South Atlantic Anomaly passages and the instrumental dead time.
With this data set,
 we extract $10^\circ \times 10^\circ$ square regions of interest (ROIs) in Galactic coordinates centered at the position of each dSph shown in Table~\ref{tb:dSph_datas}.
The dSphs in Table~\ref{tb:dSph_datas} provide DM-rich environments that enhance the DM pair annihilation rate.
The extracted data are binned into $0.1^\circ \times 0.1^\circ$ pixels and logarithmic energy bins with 8 bins per decade.
The ROI backgrounds are structured by galactic diffuse emission 
(\texttt{gll\_iem\_v07.fits} 
\footnote{ \url{https://fermi.gsfc.nasa.gov/ssc/data/access/lat/BackgroundModels.html} }),
isotropic diffuse emission
(\texttt{iso\_P8R3\_SOURCE\_V3\_v1.txt}
\footnote{ \url{https://fermi.gsfc.nasa.gov/ssc/data/access/lat/BackgroundModels.html} }),
and resolved point sources in the 4FGL-DR4 catalog
\footnote{ \texttt{gll\_psc\_v34.fit}:
\url{https://fermi.gsfc.nasa.gov/ssc/data/access/lat/14yr_catalog/}
}.

\begin{table}[H]
  \centering
  \caption{Properties of Milky Way dSphs.}
\begin{threeparttable}
  \begin{tabular}{l c c c c}
  \hline\hline
    Name & R.A. (J2000)\tnote{a} &  Decl. (J2000)\tnote{b} & Distance & $\log_{10}(J_{\rm obs})$\tnote{c} \\
         &     [deg]    &      [deg]     &   [kpc]  & [$\log_{10} ({\rm GeV}^2{\rm cm}^5)$]   \\
  \hline 
    Bootes I & 210.02 & 14.51 & 66.0 & 18.8 $\pm$ 0.22 \\
    Canes Venatici II & 194.29 & 34.32 & 160.0 &  17.9 $\pm$ 0.25 \\
    Carina &  100.41 &  -50.96 & 105.0 & 18.1 $\pm$ 0.2 \\
    Coma Berenices &  186.75 & 23.91 & 44.0 & 19.0 $\pm$ 0.25 \\    
    Draco & 260.07 & 57.92 & 76.0 & 18.8 $\pm$ 0.16 \\
    Fornax & 39.96 & -34.5 & 147.0 & 18.2 $\pm$ 0.21 \\
    Hercules &  247.77 & 12.79 & 132.0 & 18.1 $\pm$ 0.25 \\
    Leo II & 168.36 & 22.15 & 233.0 &  17.6 $\pm$ 0.18 \\
    Leo IV & 173.24 & -0.55 & 154.0 & 17.9 $\pm$ 0.28 \\
    Sculptor & 15.04 & -33.71 & 86.0 & 18.6 $\pm$ 0.18 \\
    Segue 1 & 151.75 & 16.08 & 23.0 & 19.5 $\pm$ 0.29 \\
    Sextans & 153.26 & -1.61 & 86.0 & 18.4 $\pm$ 0.27 \\
    Ursa Major I & 158.77 & 51.95 & 97.0 & 18.3 $\pm$ 0.24 \\
    Ursa Major II &  132.87 & 63.13 & 32.0 & 19.3 $\pm$ 0.28 \\
    Ursa Minor &  227.24 & 67.22 & 76.0 &  18.8 $\pm$ 0.19 \\
    Willman 1 &  162.34 & 51.05 & 38.0 &  19.1 $\pm$ 0.37 \\
  \hline\hline      
    \end{tabular}
  \label{tb:dSph_datas}
\begin{tablenotes}\footnotesize
        \item[a,b] The R.A. and the Decl. stand for Right Ascension and Declination, respectively. 
        Each coordinate refers to \url{https://ned.ipac.caltech.edu/}. 
        \item[c] J-factors derived from stellar kinematics\cite{main-referred-study2}.
\end{tablenotes}
\end{threeparttable}
\end{table}


\section{Model and Dark Matter Pair Annihilation}

We review the Higgs-portal Majorana fermion DM model~\cite{previous study}.
In the model, we introduce a gauge-singlet Majorana fermion, $\chi$, to the SM field content.
We also introduce a $Z_2$ symmetry, under which $\chi$ is odd and the SM fields are even. The relevant part of the Lagrangian involving $\chi$ reads
\bea
    {\cal L} \ \supset \ 
    \frac{i}{2}\bar{\chi}\gamma^\mu\partial_\mu\chi 
    - \frac{m_\chi}{2} \bar{\chi}\chi 
    - \frac{\bar{\chi}\chi H^\dagger H}{\Lambda} 
	\quad ,
\eea
where $H$ denotes the SM Higgs field, and $\Lambda$ is a cutoff scale of the model.
Because of the $Z_2$ parity, the neutral particle $\chi$ with mass $m_\chi$ becomes stable and provides a natural DM candidate.\footnote{
We hereafter denote the DM particle also by $\chi$.
}
The SM Higgs field $H$ comprises the physical Higgs boson $h$ and the longitudinal components of the $Z$ and $W^\pm$ bosons, denoted by $Z_L,W^\pm_L$.

As a result, the DM $\chi$ pair-annihilates into $h, Z_L, W^\pm_L$.
The latter go into cascade decays and generate $\gamma$-rays.

\begin{figure}[H]
    \centering
    \begin{subfigmatrix}{2}
    \includegraphics{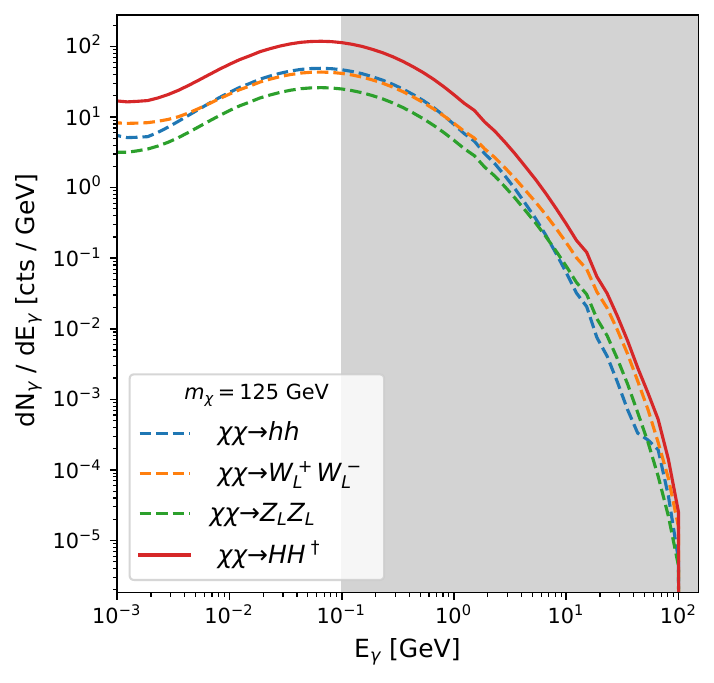}
    \includegraphics{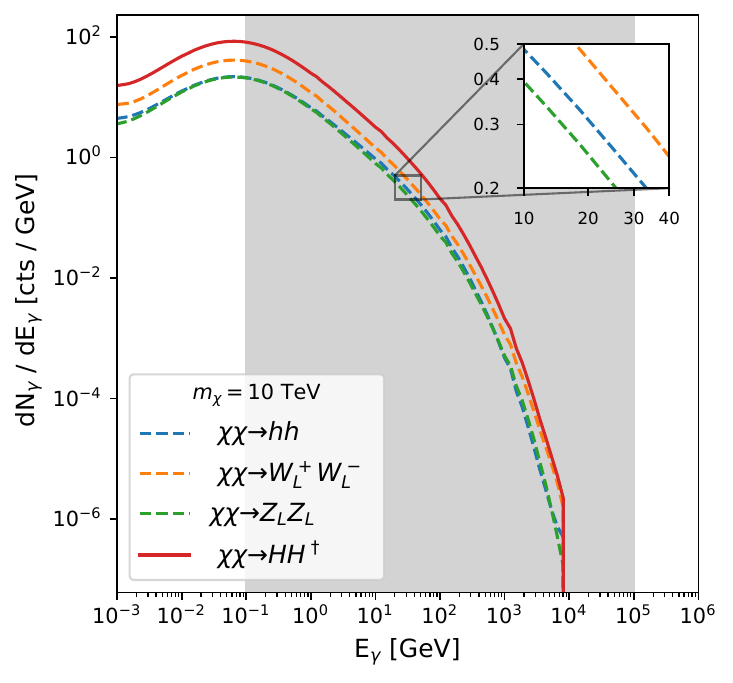}
    \includegraphics{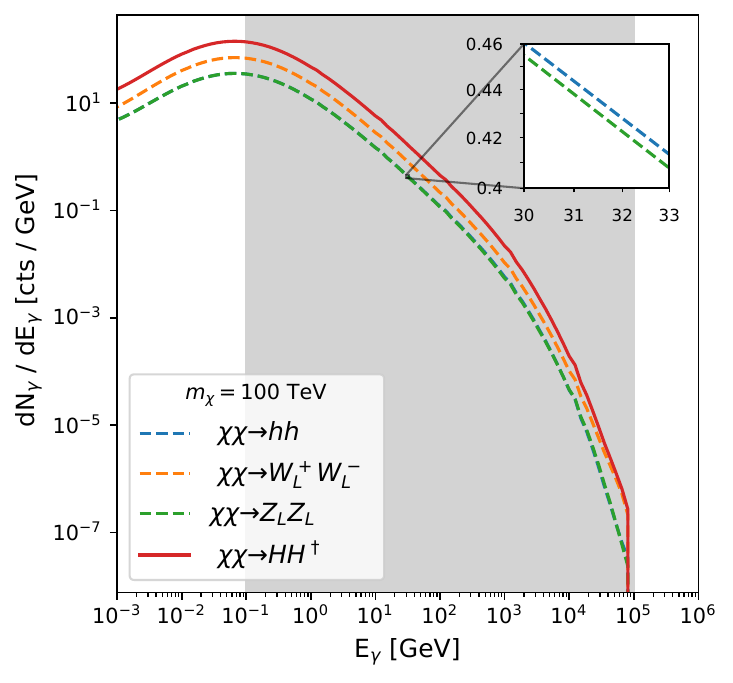}
    \end{subfigmatrix}
    \caption{Differential $\gamma$-ray yield per annihilation from the DM pair annihilation.
    In each panel, the dashed lines represent fluxes emitted from $hh$, $W^+_LW^-_L$ and $Z_LZ_L$, and the solid line is their sum.
    The gray area represents {\it Fermi}-LAT's valid energy range.
    The upper-left, upper-right and lower panels correspond to the DM mass of
    $m_\chi = 125$ GeV, 10 TeV and 100 TeV, respectively.
    }
    \label{fig:enter-labe2}
\end{figure}

The $\gamma$-ray signal flux produced from the DM pair annihilation and observed at LAT is 
\bea
    \Phi (\Delta \Omega)
     = 
    \underbrace{
    \frac{1}{4\pi} \frac{\Braket{\sigma v}}{2 m_\chi^2}\int_{E_{\rm min}}^{E_{\rm max}}\frac{{\rm d}N_\gamma}{{\rm d}E_\gamma} {\rm d}E_\gamma
     }_{ \text{particle physics} }
    \times
    \underbrace{
    \int_{\Delta \Omega}\int_{{\rm l.o.s}}\rho_{{\rm DM}}( \bm r )^2 {\rm d}l {\rm d}\Omega^\prime
    }_{ \text{J-factor} }
    \quad .
\eea
The first factor has the information about particle physics, i.e.,
 the cross section times velocity of the processes $\chi \chi \to hh, \, W_L^+ W_L^-, \, Z_LZ_L$ averaged over the DM energy distribution, $\Braket{\sigma v}$,
 the DM particle mass $m_\chi$,
 and the differential $\gamma$-ray yield per annihilation,
${\rm d}N_\gamma / {\rm d}E_\gamma $,
 integrated over the experimental energy range.
Here ${\rm d}N_\gamma / {\rm d}E_\gamma $ is computed by PPPC4 DM ID package~\cite{PPPC4}.
The package's valid mass range is between 5 GeV and 100 TeV.
In this paper,
 we explore the DM mass range from 125 GeV (Higgs mass) to 100 TeV.
Fig.~\ref{fig:enter-labe2} shows ${\rm d}N_\gamma / {\rm d}E_\gamma $ for various DM masses. 

The second factor,
 which is called J-factor,
 is not dependent on particle physics but on astrophysical information.
The J-factor is the integral of  the DM density squared,
 $\rho_{\rm DM}^2$,
 along the line-of-sight (l.o.s.) and the solid angle, $\Delta \Omega$.

\section{J-Factors for Dwarf Spheroidal Galaxies}
To compute the signal $\gamma$-ray spectrum from DM annihilation,
 we have to know the DM distribution.
Modeling the density of DM in astrophysical objects is an important topic, discussed in Refs.~\cite{HierarchicalMassModelling,Non-SphericalDarkHalos,ScalingRelations}.
The astrophysical targets we use for the analysis are dSphs of the Milky Way, 
which are approximately spheroidal object with large DM density.
We employ the following Navarro-Frenk-White (NFW) profile for the DM distribution
\cite{StructureOfColdDM_Halos,ProfileFromHierarchicalClustering}:
\begin{equation}
    \rho_{\mathrm{NFW}}=\frac{\rho_s}{r / r_s\left(1+r / r_s\right)^2}
    \qquad ,
\end{equation}
where $ r $,  $ r_s $, and $ \rho_s $ are the distance from the galaxy center,
 the NFW scale radius,
 and the NFW scale characteristic density, respectively.
To improve the accuracy of parameter estimates for the DM density profile,
 we take into account the fact that dSphs live in subhalos of the Milky Way
\cite{StructureFormation}. 
The properties of the dSphs are listed in Table~\ref{tb:dSph_datas}.

\section{ Combined likelihood analysis}

We use the {\it Fermi}-LAT data on $\gamma$-rays from the 16 dSphs, to constrain the DM pair annihilation cross section in the Higgs portal Majorana fermion DM model.

Counting experiments,
 such as {\it Fermi}-LAT observations of astrophysical $\gamma$-rays,
 commonly utilize a Poisson distribution.
We combine the Poisson likelihood functions of the 16 dSphs,
 and further add J-factor likelihood term to the Poisson likelihood to account for the statistical uncertainties on the J-factor of each dSph~\cite{previous-LAT-study1,previous-LAT-study2}.
The J-factor likelihood for target dSph $i$ is given by a log-normal distribution:
\begin{equation}
    \mathcal{L}_J\left(J_i \mid J_{\mathrm{obs}, i },\sigma_i\right)
     =\frac{1}{\ln (10) J_{\mathrm{obs}, i } \sqrt{2 \pi} \sigma_i} \times 
     \exp{ \left[ - \left(
     \frac{ \log _{10}(J_i) - \log _{10}( J_{\mathrm{obs}, i } ) }{ \sqrt 2 \sigma_i }
     \right)^2 \right]}
     \qquad ,
\end{equation}
where $ J_i $ and $J_{ {\rm obs}, i}$ are the true value of the J-factor and the measured J-factor with error $\sigma_i$, respectively. 
The combined likelihood functions for target $i$ become,
\begin{equation}
    \tilde{\mathcal{L}}_i\left(
    \mu, \boldsymbol{\theta}_i=\left\{\boldsymbol{\alpha}_i, J_i\right\} \mid \mathcal{D}_i
    \right)=
    \mathcal{L}_i\left( \mu, \boldsymbol{\theta}_i \mid \mathcal{D}_i\right) 
    \mathcal{L}_J\left(J_i \mid J_{\mathrm{obs}, i }, \sigma_i\right)
    \qquad .
\end{equation}
Here $\mu$ stands for the parameters of the DM model, namely, the DM mass $m_\chi$, and 
 the average cross section times velocity $\Braket{\sigma v}$ for the processes $\chi \chi \to hh, \, W_L^+ W_L^-, \, Z_LZ_L$.
$\boldsymbol{\theta}_i$ is the nuisance parameter set that consists of 
 the LAT analysis ($\boldsymbol{\alpha}_i$) and the dSph J-factor ($J_i$),
 and $\mathcal{D}_i$ is the LAT $\gamma$-ray data.
Then, 
 the joint likelihood function is given by
\begin{equation}
    \mathcal{L}\left( \mu, \boldsymbol{\theta} \mid \mathcal{D}\right)
     =
     \prod_i \Tilde{ \mathcal{L} }_i\left( \mu, \boldsymbol{\theta}_i \mid \mathcal{D}_i\right)
     \qquad .
\end{equation}
The test statistics (TS) for $\mu$ is defined as
\begin{equation}
    \operatorname{TS}( \mu)
     = 
     -2 \ln \left(\frac{\mathcal{L}(\mu, \hat{\boldsymbol{\lambda}})}{\mathcal{L}(\hat{\boldsymbol{\theta}})}\right)
     \qquad ,
\end{equation}
 where $\hat{ \boldsymbol{\theta} }$ are the values that maximize $\mathcal L$ globally,
 and $\hat{ \boldsymbol{\lambda} } $ are the values that maximize ${\mathcal L}$ for a given DM mass and $\Braket{\sigma v} $.
The TS shows an approximate $ \chi^2 $ distribution and the 95$\%$ confidence level (CL) limit on $\mu$ is obtained from the relation TS = 2.71.

We use the profile likelihood method~\cite{LimitsAndConfidenceIntervalsInThePresenceOfNuisanceParameters,TASI_2023_IndirectDetection}
 to derive upper bounds on the average cross section times velocity $\Braket{\sigma v}$ for various values of the DM mass $m_\chi$.

\section{Result}
\begin{figure}[H]
   \centering
       \includegraphics{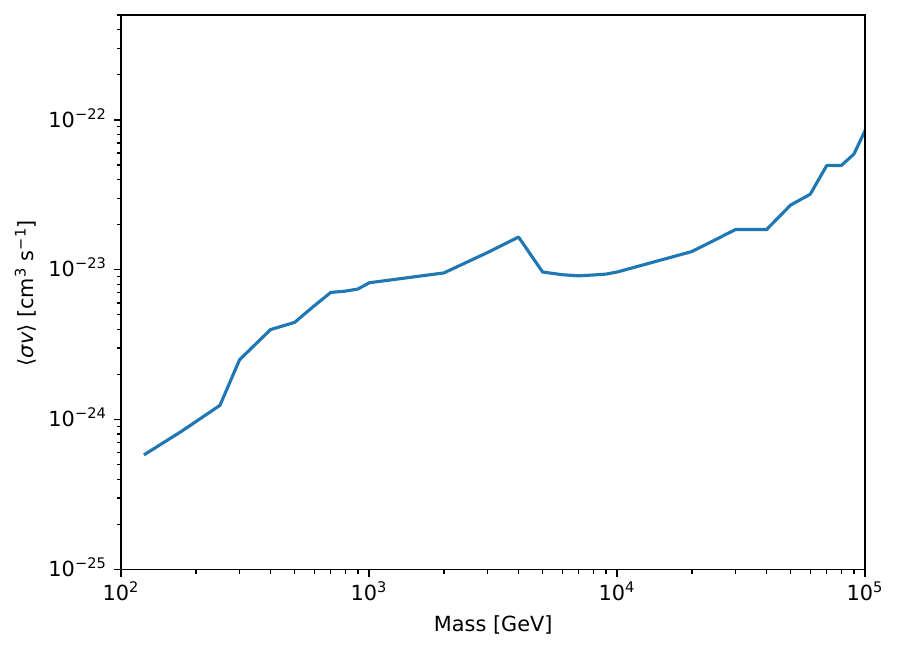}
    \caption{
    $95\%$ CL upper limits on the average DM pair annihilation cross section times velocity $\Braket{\sigma v}$ in the Higgs portal Majorana fermion DM model, for the DM mass between 125 GeV and 100 TeV,
 obtained from the {\it Fermi}-LAT data on $\gamma$-rays from the 16 dSphs.
    }
    \label{fig:enter-labe3}
\end{figure}

In Fig.~\ref{fig:enter-labe3},
 we show the $95\%$ CL upper limits on the average DM pair annihilation cross section times velocity $\Braket{\sigma v}$.

\begin{figure}[H]
  \begin{subfigmatrix}{2}
    \subfigure[]{\includegraphics[scale=1]{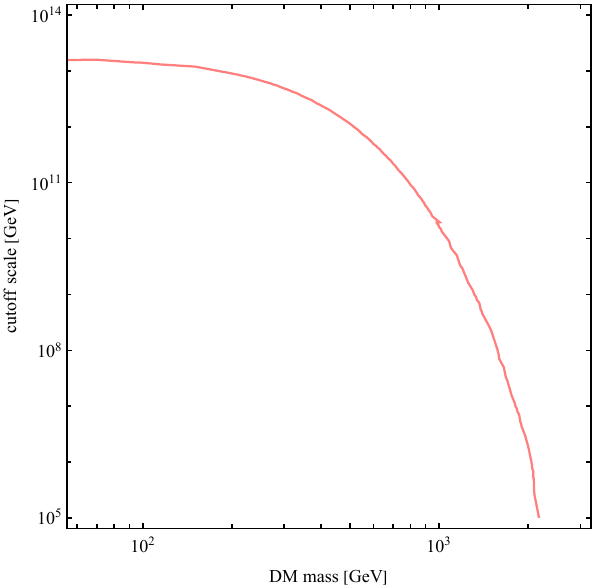} }
    \subfigure[]{\includegraphics[scale=1]{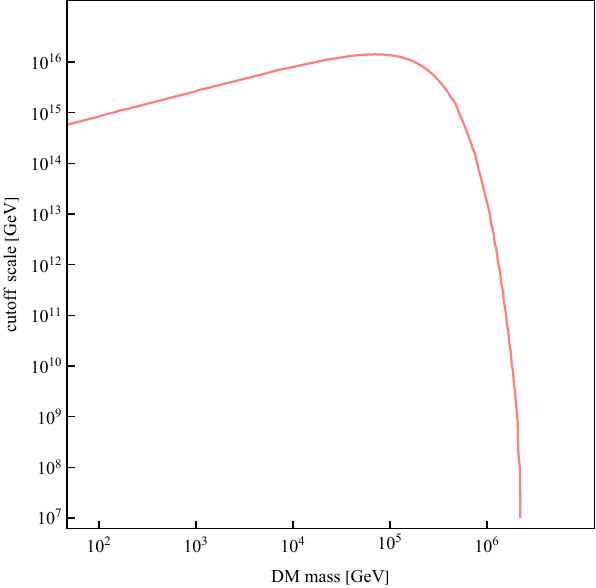} }
  \end{subfigmatrix}  
  \caption{
The relation between the cut-off scale $\Lambda$ and the DM mass $m_\chi$ for fixed reheating temperature $T_{\rm RH}$
  that reproduce the correct DM relic abundance $\Omega_{\rm DM} h^2 = 0.12$. 
The left panel is for $T_{RH}=100$ GeV and the right panel is for $T_{RH}=100 $ TeV.
}
  \label{fig:Contourplot}
\end{figure}

\begin{figure}[t]
  \begin{subfigmatrix}{2}
    \subfigure[]{\includegraphics[scale=1]{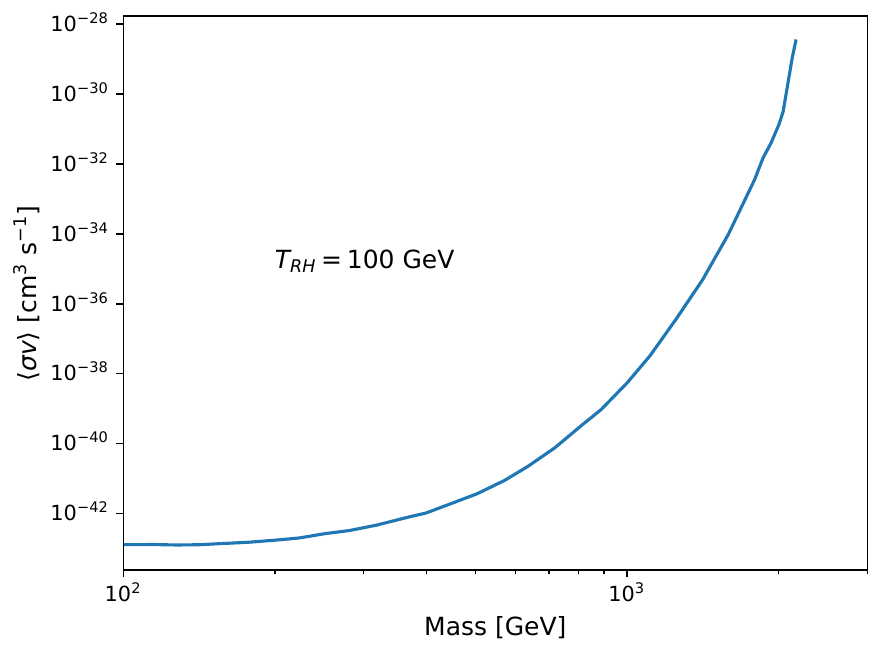} }
    \subfigure[]{\includegraphics[scale=1]{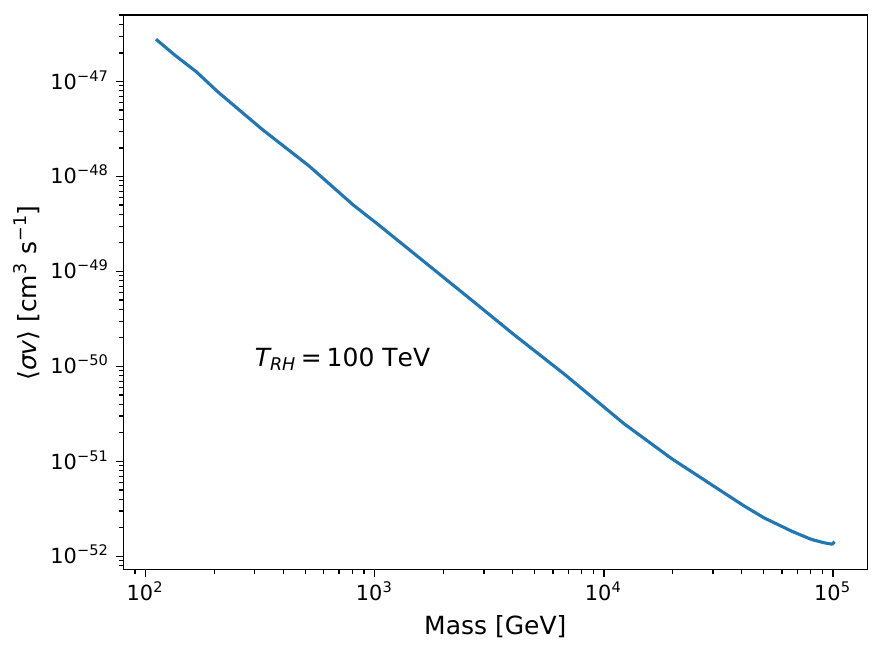} }
  \end{subfigmatrix}  
  \caption{
  $\Braket{\sigma v}$ for fixed $T_{\rm RH}$ and for $(m_\chi, \Lambda)$ that reproduce the observed DM relic abundance $\Omega_{\rm DM} h^2 = 0.12$, given in Fig. \ref{fig:Contourplot}.
  The left panel is for $T_{\rm RH} = 100 $ GeV and the right panel is for $T_{\rm RH} = 100$ TeV.
}
  \label{fig:sigmavInPriorStudy}
\end{figure}

We compare Fig.~\ref{fig:enter-labe3} with the result of Ref.~\cite{previous study} that 
 is the relation between the cut-off scale $\Lambda$ and the DM mass $m_\chi$ that reproduce the correct DM relic abundance $\Omega_{\rm DM} h^2 = 0.12$,
 for a given reheating temperature $T_{\rm RH}$.
This relation is reviewed in Fig.~\ref{fig:Contourplot} for $T_{RH}=100$ GeV and $100 $ TeV.
Using $(m_\chi, \Lambda)$ values in Fig.\ref{fig:Contourplot},
 we calculate the average DM pair annihilation cross section times velocity $\Braket{\sigma v}$.
In calculating $\Braket{\sigma v}$, the distribution of the energies of a DM pair {\it at present} is approximated as follows
(we write the energies of a DM pair as $E_1,E_2$):
If DMs are most abundantly produced (via freeze-in mechanism) when they are non-relativistic,
 the distribution of a DM pair {\it at present} is proportional to $e^{-(E_1+E_2)/T_a}$
 with $T_a$ being the temperature at the time of the most abundant production.
This is because, when a DM pair is created, the sum of their energies is equal to the sum of the energies of the SM particles from which the DM pair is created,
 and these SM particles are in thermal equilibrium and follow the Boltzmann distribution.
After the DM pair is created, their energy distribution 'freezes' because they are non-relativistic and is preserved until present.
If DMs are most abundantly produced when they are relativistic,
 the distribution of a DM pair {\it at present} is proportional to $e^{-(E_1+E_2)/T_\chi}$ where $T_\chi \sim m_\chi$.
This is because the energy of a DM decreases with the temperature of the radiation until the DM becomes non-relativistic, although the DM is not in thermal equilibrium.
After it becomes non-relativistic, its energy distribution 'freezes' and is preserved until present.
Given the above approximation, $\Braket{\sigma v}$ is calculated as
\bea
	\Braket{ \sigma v }
		= \frac{g_\chi^2}{64 \pi^4} T \frac{1}{ n_{\rm eq}^2(T) } 
				\int^\infty_{4m^2_\chi} ds ~s \sqrt{s-4m^2_\chi}  (\sigma v) K_1\left( \frac{\sqrt s}{T}  \right)
	\quad ,
 \label{eq:sigmav}
\eea
where $T=T_a$ or $T_\chi$ depending on whether DMs are most abundantly produced when they are non-relativistic or relativistic.
Here $g_\chi$=2 counts the physical degree of freedom of the Majorana fermion $\chi$,
 $n_{\rm eq}(T)$ is the number density for the Boltzmann distribution with temperature $T$,
 $(\sigma v)=W^{ {\rm 2-body } }_{ {\rm ij } } / s$, 
 where $W^{ {\rm 2-body } }_{ {\rm ij } }$ is the (unpolarized) annihilation rate per unit volume corresponding to the covariant normalization of $2E$ colliding particles per unit volume and $s$ is the invariant mass,
 and $K_1$ is the modified Bessel function of the second kind of order 1.
$\Braket{\sigma v}$ calculated as above, for $(m_\chi, \Lambda)$ that reproduce the correct DM relic abundance,
 is presented in Fig.~\ref{fig:sigmavInPriorStudy},
 for $T_{\rm RH}=100$~GeV and 100~TeV.

Comparing Fig.~\ref{fig:sigmavInPriorStudy} with Fig.~\ref{fig:enter-labe3}, we find that
 for $T_{\rm RH} = 100$ GeV and for $m_\chi = 2100$ GeV, the current {\it Fermi}-LAT data constrain $\Braket{\sigma v}$ to the value $10^5$ times larger than the value 
 that reproduces the correct DM relic abundance.
Note that 2100 GeV is the upper bound of $m_\chi$ when $T_{\rm RH} = 100$ GeV, which has been obtained in our precious work~\cite{previous study}.
For smaller DM masses and/or for $T_{\rm RH} = 100$ TeV, the constraint from the {\it Fermi}-LAT data, in comparison to the value of $\Braket{\sigma v}$
 that matches the correct DM relic abundance, is much weaker.
\\

\section{Conclusion}

We have investigated the $\gamma$-ray signal from DM pair annihilation in the Higgs portal Majorana fermion DM model.
In the model, the pair annihilation of DMs through a Higgs portal higher-dimensional operator creates a pair of the Higgs bosons or the longitudinal modes of $W,Z$ gauge bosons,
 which then go into cascade decays and generate signal $\gamma$-rays.
We have performed a likelihood analysis using the {\it Fermi}-LAT data on $\gamma$-rays from the 16 dSphs of Table~\ref{tb:dSph_datas},
 and obtained upper bounds on the average DM pair annihilation cross section times velocity $\Braket{\sigma v}$ in Fig.~\ref{fig:enter-labe3}.
These bounds are compared with the values of $\Braket{\sigma v}$ that correspond to the sets of the DM mass and the cut-off scale that reproduce the correct DM relic abundance.
We have found that when the reheating temperature is $T_{\rm RH} = 100$ TeV and the DM mass is $m_\chi = 2100$ GeV,
 the {\it Fermi}-LAT data constrain $\Braket{\sigma v}$ to the value $10^5$ times larger than the value corresponding to the correct DM relic abundance.
The ratio of the upper bounds from the {\it Fermi}-LAT data over the values of $\Braket{\sigma v}$ that reproduce the correct DM relic abundance, is much smaller 
 for smaller DM masses and/or for $T_{\rm RH} = 100$ TeV.
\\

\section*{Acknowledgments}
This work is partially supported by Scientific Grants by the Ministry of Education,
Culture, Sports, Science and Technology of Japan, No. 21H00076 (NH) and No. 19K147101 (TY).


\end{document}